\documentclass[prd,nofootinbib,preprintnumbers,
]{revtex4}
\usepackage{graphicx,epsf,epsfig}
\usepackage{amsmath, amssymb, amsfonts, bbm}
\usepackage{amssymb}
\usepackage{tabls,multirow}

\def\vev#1{\left\langle #1\right\rangle}

\def\hbar{\hspace{0pt}\raisebox{1pt}{$-$} \hspace{-7pt} h}

\newcommand{\be}{\begin{equation}}
\newcommand{\ee}{\end{equation}}
\newcommand{\bd}{\begin{displaymath}}
\newcommand{\ed}{\end{displaymath}}
\newcommand{\bea}{\begin{eqnarray}}
\newcommand{\eea}{\end{eqnarray}}
\newcommand{\nn}{\nonumber}

\newcommand{\ta}{\tilde\alpha}
\newcommand{\tb}{\tilde\beta}

\newcommand{\da}{\dot\alpha}
\newcommand{\db}{\dot\beta}

\newcommand{\e}{\eta}
\newcommand{\nnu}{\nonumber\\}

\newcommand{\oot}{\overline {126}}

\def\so10{$SO(10)$}

\begin{document}
\title{  Babel  on the  Petaplex site: On Rival  Calculational methods   in SO(10) MSGUTs}
\date{\today}

\author{ Charanjit S. Aulakh}\email{ aulakh@pu.ac.in}
\affiliation{ Dept. of Physics, Panjab University,Chandigarh,
INDIA 160014}\hfil\break

 \affiliation{Indian Institute of Science, Education and Research
(Mohali),
 Sector 26, Chandigarh 160019  }

\begin{abstract}
We compare and contrast the computations that lead to the NMSGUT
spectra and  Yukawa couplings that appeared\cite{nmsgut} in 2006
and a recent recalculation of the same\cite{malinsky}. We  argue
that an  explicit component based   method of computation
jeopardizes the power of SO(10) and its sub-groups to organize, in
a unified and automatically  phase correlated way, computations of
dynamics beyond the basic mass matrix computation. The correct
(one line) prescription   for  generating MSSM Yukawas from SO(10)
ones was given  in \cite{ag2} and requires no computation beyond
the identification of null vectors of the Higgs doublet mass
matrix and the Clebsches   given in \cite{ag1,ag2}. It was already
used to derive all fermion Yukawas and Majorana masses in
\cite{ag2,nmsgut}. We thus urge the adoption of a uniform
 notation and methodology based on descent from SO(10) to the SM through
 the Pati-Salam  maximal subgroup of SO(10)  to  avoid Babel in
this rapidly developing and highly promising subject.

 \end{abstract}
\maketitle
\section{Introduction}

Since the discovery of neutrino mass in the late nineties, SO(10)
GUTs and particularly Supersymmetric SO(10) GUTs have become the
leading contenders for unification.  We had
constructed\cite{mslrms} the (renormalizable and
non-renormalizable)  Minimal Supersymmetric Left Right
Models(MSLRMs), which have generically high scale B-L symmetry
breaking, automatic R-parity as a part of the gauge group, and
naturally accommodate both Type I and Type II seesaw mechanisms
for neutrino mass. We thus received the news of neutrino
oscillations from SuperKamiokande\cite{superK} with some glee
since the Seesaw estimate for the  ${B-L}$ breaking scale
$M_{B-L}$  corresponding to the neutrino masses indicated by
\cite{superK} was so large ($ \geq 10^{14}~ GeV$). Thus  the
construction of Supersymmetric SO(10) GUTs incorporating the
insights from our study of MSLRMs was naturally high on our
agenda\cite{genealog} and a model based on the $\mathbf{45 \oplus
54 \oplus 126\oplus \oot}$ Higgs system was duly constructed by
us\cite{abmrs01}. Nevertheless having long been aware\cite{aulmoh}
of the formidable calculational problems in handling SO(10) group
theory, particularly the translation from orthogonal group spinor
labels to unitary group labels, we commenced, in 2000, development
of a systematic decomposition of SO(10) labels and invariants into
those of  its `Pati Salam' maximal  subgroup : $SU(4)\times
SU(2)_L\times SU(2)_R$. The other maximal subgroup, $SU(5)\times
U(1)$, not accidentally, also received attention contemporaneously
for essentially the same purpose\cite{nathsyed}.
 The results from these calculations \cite{ag1,nathsyed},
have furnished manuals for handling any conceivable invariant
decomposition in SO(10) into unitary subgroup invariants that is
available to all workers in the field. Our intent was that the
uniformized methods and notations should make communication and
comparison of the   complex expressions obtained when decomposing
SO(10) labels into Unitary group labels easy, and to provide
`Clebsches' that were otherwise hard to compute such as those for
the spinorial 16-plet representation.  Unfortunately, just as in
the biblical story, life has turned out to be more complex and
less innocent than we might, in our naivete,  have believed
earlier.

Firstly just as we completed the `SO(10) a la Pati Salam'
decomposition methodology\cite{ag1}, having thoroughly appreciated
the complexity we had uncovered, we searched\cite{genealog} for a
model even simpler than the one we had analyzed\cite{abmrs01}. We
soon realized\cite{abmsv}  that the old model\cite{aulmoh,ckn} (
which we named as the Minimal Supersymmetric Grand Unified
Theory(MSGUT))  studied right at the beginning of the Susy GUT era
was the best and simplest or\emph{ minimal} home for the ideas on
  R-parity and Susy LR subsuming  GUTs that we had
  developed\cite{mslrms,abmrs01}. A very
significant breakthrough was achieved in that the spontaneous
symmetry breaking equations could be reduced\cite{abmsv} to the
study of a single cubic equation with a single coupling parameter
making the problem completely soluble. We therefore immediately
applied\cite{ag1} our newly developed methods to analyze both the
symmetry breaking at high scale and the resultant spectra. Indeed
already the second version, in August 2003 \cite{ag1}, of our
manual on SO(10) decomposition (which we delayed sending to press
till 2005 in order that it retain as few calculational mistakes
and typos as possible) contained the Clebsches of the matter
fermion couplings to the 3 Fermion Mass(FM)  Higgs representations
in SO(10) (the possible yukawa couplings are just :
 $\mathbf{16.16.(10\oplus120\oplus\oot)}$), as well as the mass
matrices of the two most important superheavy particle types : the
MSSM Higgs type doublets and the Higgs triplets
$[3,1,\pm{\frac{2}{3}}]$ responsible for proton decay. We also
used these to derive the effective potential for $d=5$ operator
mediated proton decay\cite{ag1}.

 Clearly the calculation of the complete spectrum
and the computation of GUT exotic effects was the next logical
step since that was just the purpose of the whole development of
calculational methods.  We duly proceeded with these calculations,
unfortunately delayed somewhat by personal circumstances, and
little aware that our leisurely pace was about be forced,
completed it finally(along with RG analysis of threshold
corrections) only in May 2004. Meanwhile the excitement and
attention raised by our revival of the old model
of\cite{aulmoh,ckn} motivated two groups, one composed of our own
collaborators\cite{bmsv} and another comprising\cite{fuku} one of
the developers of an old method for computing mass matrices in
GUTs\cite{heme} and their newly attracted collaborators. Both
groups adopted the method of\cite{heme} in preference to our
systematic and general method (which, although complete and
general  requires considerable patience and effort, while the
other method can be implemented on a computer algebra system and
thus allows more rapid computation of mass matrices : at the price
of abandoning the standard field theoretic and tensorial methods
which can also handle  more general questions such as the
decomposition of all  SO(10)   \emph{interactions} into unitary(PS
or SM) labels). Moreover the necessity of making lists of SO(10)
representation component-wise phase choices that were near
impossible to code in simple generative rules that could be
communicated  easily   made it impossible for us to agree with
such shortsighted adoption of a limited and opaque technology in
preference to the almost transparently  luminous(for us!)
symmetries of the PS decomposition of SO(10) that we had so
laboriously developed : but which, unfortunately, appear to be
less than `penetrable' to some. On the other hand perhaps these
proprietary conventions grew for the same strong reasons that
Babel and private property always threaten human cooperation. In
any case with the adoption of two methods(and three sets of phase
conventions) the rosy dream of a uniform notation rapidly faded in
the face of the harsh reality of a struggle for citation priority
and allegations of incorrect results in preference to cooperation
to pin down the inevitable errors by cross checking. These issues
are discussed at length in \cite{consistency}.

Due to the great interest in whether GUTs can account for the
fermion mass and mixing data  and the promise of SO(10) in this
regard, MSGUTs attracted a great deal of attention focussed on
these questions\cite{allferm}. Since then, after very promising
developments as regards the generic fermion fitting properties,
there have been some dramatic reversals\cite{gmblm,blmdm}  and
then further development\cite{blmdm,nmsgut,pinmsgut,nmsgutII} of
the ability to fit all fermion data using formulae specific to the
MSGUTs. MSGUTs  have thus matured to the point of becoming  quite
fully specified falsifiable theories facing challenges (which so
far they have, in one avtar or another, duly overcome) which
promise to put the surviving version  to the stress tests that
will either certify its health or consign it to the dustbin of
history. Indeed, using the very spectrum, Clebsch and Yukawa
couplings that we had computed in \cite{ag1,ag2}  we were able to
show\cite{gmblm,blmdm} that the generic scenarios of fermion
spectrum fits that had been shown to be feasible\cite{allferm}
were in fact \emph{not} feasible in the fully specified MSGUT
since the neutrino masses attained, whether by the Type I or Type
II seesaw mechanisms, were far too small. Interestingly a group,
including the author of \cite{malinsky} working at Trieste,
confirmed our results(announced at PLANCK05, held in Trieste in
June 2005,  and again in December 2005\cite{blmdm}) in April
2006\cite{bertmal}.

    In \cite{blmdm} we also  pinpointed  the
reasons for this failure as the necessary largeness of the
$\mathbf{\oot}$ couplings (due to their dual function : they
 generate neutrino Majorana masses  and
make CKM mixing in the effective MSSM Yukawa couplings possible).
 So we proposed\cite{blmdm,nmsgut} a new version (which we called the Next or New MSGUT (NMSGUT))
 in which   the theory was completed by inclusion of the
remaining possible FM Higgs,  namely the $\mathbf{120}-$plet
(which has couplings antisymmetric in
 family indices which are well adapted to generating CKM mixing
 angles) and a re-assignment of roles  in which the $\mathbf{\oot}$
 couplings,   become  very small and thus play little or no role
 in charged fermion masses, but can then   boost the Type I
 seesaw mass since small $\mathbf{\oot}$ couplings lower the
 righthanded neutrino masses. To implement these ideas we made some
 perturbative studies\cite{core,msgreb} and more importantly
 already in\cite{blmdm} and completely in \cite{nmsgut} used our SO(10)
 decomposition technology\cite{ag1}  to compute the same
 information for the NMSGUT: fermion Yukawa couplings, superheavy mass matrices,
 Baryon violation effective superpotentials
 and superheavy  threshold effects in the RG flow  that
we had earlier computed for the MSGUT\cite{ag1,abmsv,ag2}. This
paper was released in December 2006 and by now we are already
using its results in large scale computer studies of realistic
fitting of all spectra in
 NMSGUTs\cite{precthresh,pinmsgut,nmsgutII} which show that indeed
our proposal is effective in generating viable and NMSGUT specific
fits to all fermion data, neatly evading all constraints found
earlier. For this they need to invoke the participation of
threshold corrections at $M_S$  and then also furnish information
on the sfermion spectra compatible with fermion masses, and thus
opened up the entire spectrum for theoretical investigation vis a
vis falsifiability.

 Very recently, just as we released\cite{nmsgutII} at a
conference at ICTP, Trieste\cite{nmsgut@cpt@ictp}  we encountered
the author of\cite{malinsky}, who shortly thereafter released his
recalculation of the results of the appendices A and C of
\cite{nmsgut}. While, as we emphasized above, we have always
welcomed the opportunity to cross check our results against
independent computations  to uncover any discrepancies and errors
of detail, it is unfortunately true, as explained in detail in
\cite{consistency}, that it is difficult to compare computations
which use   the component based approach of \cite{fuku,bmsv} and
our own Lagrangian oriented SO(10)/PS tensorial approach
 up to  the point where phase differences can be
cross checked. The reasons for this are fully explained in
\cite{consistency}.

The author of \cite{malinsky}  opined that the phase structure of
our analysis was `impenetrable'. While, in the absence of detailed
reasons, such a judgement is entirely subjective    it
sufficiently motivated us to study the paper in order to evaluate
the validity  of the claims. We found that the author's method
shared the non-generality of the component based result as used
earlier and even made  errors in the actual phase specification.
 Due to the
importance of the basic framework for clarity in this complex
subject,  we thought it behooved us to point out once again the
virtues of our method and the simple and correct prescription to
determine the fermion Yukawas. We do this in the hope  of damping
down the Babel that has developed in the SO(10) PetaGeV-tower
construction that we have spent  much time and effort to lay the
foundations of
\cite{aulmoh,mslrms,abmrs01,ag1,ag2,abmsv,gmblm,blmdm,nmsgut,core,msgreb,precthresh,pinmsgut,nmsgutII}.

In this comment we therefore first discuss in Section 2 the
contents of the two papers \cite{nmsgut,malinsky} to evaluate
their relative overlap.  In Section 3. we compare and evaluate the
component method for fixing phases adopted in works of this
type\cite{bmsv,fuku,malinsky} relative to our own tensorial
method.  We then  evaluate the `penetrable', `optically' optimal
etc method\cite{malinsky} of obtaining MSSM Yukawa couplings from
NMSGUT ones and  point out some errors it arrives at. We feel
these might have been easier to detect in a compact  tensor
notation like ours  rather than an  extended explicit naming of
fields and components that inhibits conceptual clarity and hides
the elementary properties of the basis changes.

\section{Comparison of paper contents }

 Before proceeding let us first see the relation to our
previous work, and the notice he had of it, that author himself
gives in a Note Added at the end of his paper\cite{malinsky}.

``A day before finishing this manuscript the author's attention
was drawn to the preprint \cite{nmsgut} where the relevant part of
the next-to minimal SUSY $SO(10)$ model has been previously
studied from a similar perspective. As far as one can see through
the jungle of different notation, normalization and conventions
the results therein seem to agree with those given in this study,
\textbf{\emph{but especially the phase conventions that we have
spent so much time on arguing about their importance here are
virtually impenetrable in \cite{nmsgut}}}. Moreover, since the
method we employed was completely different I believe that the
current study is worth and does indeed provide a valuable and
entirely independent survey of many of the crucial and technically
rather demanding prerequisites of any numerical analysis of the
NMSGUT. Despite from that, there could still be a good case for
even a further check of ours as well as Aulakh \& Garg's results,
in particular when it comes to the phases and matching(s).
\textbf{\emph{Apart from all this, the current study is certainly
much more detailed}} and we pushed hard to make it maximally
self-contained so that a careful and patient reader should be able
to read and potentially reproduce all the results with just the
ingredients given here and in the 'canonical' MSGUT reference
\cite{bmsv}. \textbf{\emph{On top of that, a lot of extra
information provided in Sections(IV)Higgs Sector  Mass Matrices,
VI)NMSGUT Yukawasector and in particular in Sections V) Goldstones
and in Appendix  } does not have any counterpart in }
\cite{nmsgut}."

The two principal emphases to note-which we have highlighted- here
are
\begin{itemize}

\item The author feels that his phase conventions in comparison to
ours are `penetrable' because he has spent so much time arguing
about their importance. This is like saying that the way to the
 the front door  is long indeed because you went via your window and
  the moon ! As we shall
see in detail our   phase conventions for   every level of descent
down from SO(10)  to the  SM gauge group  are specified at one
shot  when we  decompose\cite{ag1} the fundamental vector(10-plet)
and spinor(16-plet) irreps (from which two all irreps of Spin(10)
can be obtained- with fixed phase and normalization conventions-
by tensor products) through the PS maximal subgroup down to the
standard model (with  the standard embedding of the SM in the PS
group). \textbf{\emph{Once this is done there is simply no phase
ambiguity left ! }}  Thus we provide\cite{ag1} the technology for
translating SO(10) tensors to PS and SM ones and explicitly
provide decompositions(i.e all Clebsches but in a field theoretic
notation adapted to working with SO(10) lagrangians) of all the
interactions in the theory that are not trivial to write down.
This \emph{generative and prescriptive method } is to our mind the
only efficient and modern method of proceeding  : not(see below) a
list of phases for more than 100 individual states occupying no
less than 11 pages appended with the facile assertion that anyone
with sufficient determination and time can check and generate the
rest for himself by applying colour and electroweak gauge
transformations on the explicit states ! Why would one set oneself
such a headache if a tensor method is available to resolve all
group theoretic invariants ? Moreover   although the author of
\cite{malinsky} `` spent so much time on arguing about their
importance ''  his phase conventions   in fact contained an
manifest error (the relation between GUT and MSSM fields did not
maintain holomorphicity) which terminated our comparison.

\item   The author also claims that \textbf{\emph{``the current
study is certainly much more detailed ...... on top of that, a lot
of extra information provided in Sections IV)Higgs Sector Mass
Matrices, VI)NMSGUT Yukawasector and in particular in Sections V)
Goldstones and in Appendix    does not have any counterpart in
}}"\cite{nmsgut}.

\end{itemize}
 These statements  are quite unfounded. Let us therefore
 list for comparison what our 61 page
paper\cite{nmsgut} actually contains besides the obvious common
starting points  :

 \begin{itemize}

\item{Section \textbf{2.1.1}} :  A discussion of the
charateristics of the GUT SSB solutions in terms of the three
branches of the cubic.

\item{Section \textbf{2.2} and \textbf{3}} :  Group decomposition
of the $\mathbf{120}$ and a
  \textbf{\emph{ decomposition of the complete additional
superpotential  into PS sub-invariants i.e the explicit clebsches
for SO(10) to PS  }} (the MSGUT part was already decomposed
in\cite{ag2}). Using these  decompositions  it is trivial to read
off not only the mass matrices but also the superpotential
couplings  between any SM fields in the theory. The latter has no
counterpart in \cite{malinsky} and cannot even be sensibly stated
in that notation.

\item{Section \textbf{3.1, 3.2, 3.3 }, \textbf{Appendix A}} : The
explicit SM field multiplets which acquire Unmixed chiral, Mixed
pure chiral and Mixed Chiral-Gauge (including a description of the
goldstone substructure in terms of null eigenvectors) mass terms
and the explicit list of mass matrices (Appendix A) . It is worth
emphasizing that our mass matrices have rows and columns labelled
by SM \textbf{\emph{tensor multiplets}} so that they directly give
the mass terms in the Superpotential with correct phases and
contractions for all fields without any need to refer to extensive
pages of individual component phase  assignments  (not to mention
gauge transformations of the individual  states  to generate the
rest of the (500 ) states and \emph{their}  phases).

\item{Section\textbf{ 4.}} This is an extensive analysis  of the
Renormalization Group threshold corrections due precisely to the
spectra calculated and their physically highly significant
implications (such as a generically raised unification scale over
the viable regions of the parameter space). This is a major
physics component of \cite{nmsgut} with no counterpart
in\cite{malinsky}.

\item{Section \textbf{5.1}} :  MSSM Fermion Mass Formulae and
Yukawa couplings in terms of GUT couplings.  This set of formulae
which includes all Fermion mass relevant clebsches \cite{ag1} and
the neutrino masses was first derived correctly for both the
MSGUT\cite{ag2,gmblm} and the NMSGUT\cite{blmdm,nmsgut} by us in
terms of the null eigenvectors(after fine tuning) of the Higgs
mass matrices computed by us in\cite{ag1,blmdm}. We emphasize that
since the Clebsches of the coupling of FM Higgs to spinors could
not be computed without a method for decomposition of the cubic
invariants $\mathbf{16.16.(10\oplus \oot\oplus 120)} $ it is
necessary to choose and assign phases and normalizations in a
coordinated way and this was indeed the motivation for our
decomposition of SO(10) a la Pati salam\cite{ag1} where simple,
complete and generative rules (not lists of phases and
normalizations for individual components of large irreps) for
fixing phases and normalizations were given.  Indeed the these
very Clebsches were in fact used by
 subsequent  papers on MSGUTs e.g \cite{bmsv0511} and others;
  not always   with correct and proper attribution.

 These formulae were  then  used to show that neutrino masses calculated using MSGUT
formulae were too small\cite{gmblm}. These results were reported
already in May 2005 at PLANCK05 where the author of
\cite{malinsky} was in the audience. In fact he wrote a
paper\cite{bertmal} checking our result of too small neutrino
masses in the MSGUT  a year later after we had already proposed
the NMSGUT structure\cite{blmdm}.  As we shall explain below  the
confusion in \cite{malinsky} concerning the correct procedure to
connect NMSGUT to MSSM Yukawas   can  be understood in terms of
the fact that he was persistently using legacy, but incorrect,
unitary transformation formulae (between the original GUT basis
and the mass diagonal basis) that violate the basic structure of
the theory. We gave  the correct simple one line rule for
calculating these coefficients  as long ago as
2004\cite{ag2,gmblm,blmdm}.    In our opinion it is because  the
structure of the transformation rules was obscured by the unwieldy
explicit and obscure notation   that  the violation of
holomorphicity by incorrectly set up transformation formulae was
missed earlier by both authors and readers.

\item{Sections \textbf{5.1, 5.2,5.3,5.4 }:} Discussed CP violation
and the vital question of the actual fits of fermion data as it
stood at that time and the indications and contradictions between
different approaches : which have recently all been resolved and
unified\cite{pinmsgut,nmsgutII}. No counterpart in
\cite{malinsky}.

\item{Section \textbf{6}} : In this section we completed the
Baryon decay effective superpotential (dimension 4) that we
derived for the MSGUT   already in\cite{ag1,ag2} by adding the new
terms due to the $120-$plet and discussed the unitary
transformation to  MSSM basis sets. No counterpart in
\cite{malinsky}.

\item{Appendix \textbf{B}} :  In Appendix B we did a SU(5)
assembly recheck of our PS decompositions and showed that indeed
the MSSM multiplets and their mass terms could be reorganized
according to the other maximal subgroup when the superheavy vevs
were such that the unbroken symmetry included SU(5). No
counterpart in \cite{malinsky}

\item{Appendix \textbf{C}} :  Finally and very importantly for
later calculations\cite{nmsgutII} we gave explicit expressions for
the null eigenvectors of the Higgs doublets mass matrix which are
the crucial ingredient in defining the effective MSSM and its
Yukawa couplings. While the `weights' given in \cite{malinsky} may
turn out to be the same if the relation between the long lists of
component-wise phase conventions in\cite{malinsky} and the
\emph{rules} given in \cite{ag1} is worked out, we shall show that
their interpretation is  facilitated by our compact notation.

\end{itemize}

Thus  the computation of \cite{malinsky} covers only the
repetition of the mass matrix calculations in an idiosyncratic
phase convention i.e our Sections \textbf{3.1,3.2,3.3} and
Appendix \textbf{A}, and an attempt at reproducing our Section
\textbf{5} which seems flawed. We fail to see in what sense it
could be judged more complete or what new relevant information was
added. From the above listing the reader may judge for herself
which computation is more complete; not to speak of correct.

\section{Comparison of Methods and Results}

The crux of the method used in\cite{malinsky} and antecedent
papers can be appreciated by first quoting from \cite{malinsky}
what he sees as the determining rationale of his method (our
italics and our text-compactifying interpolations in square
brackets): ``For sake of illustration let us remark that there is
in total 13,321,010 terms in the sums in [the NMSGUT
Superpotential] (out of which 2,111 thousand terms come from the
new piece $W_{\rm H}^{\bf 120}$), but fortunately 'only' 1,190,170
of them are non-zero by antisymmetry of the tensors under
consideration ($W_{\rm H}^{\bf 120}$ then accounts for 338,400 out
of this number). Thus, perhaps the only reasonable strategy of
handling all these contributions is to work with the
antisymmetrized combinations rather than with the very components
of the antisymmetric tensors ....  "

``In what follows we shall pass through the whole plethora of
 the Higgs sector states and write down the corresponding mass
 (fermionic) matrix for each subspace corresponding to a set of
  fixed values of  [its Casimirs and SM quantum numbers]
   \textbf{\emph{\emph{choosing a single
representative configuration of the Cartan eigenvalues for each
value of the relevant Casimir}}},...... .  {The mass matrices for
all the other components with the same[ Casimirs and hypercharge
but different colour or electroweak weights] \textbf{\emph{can be
(if desired) obtained in a straightforward manner by the relevant
$SU(3)_{c}$ and/or $SU(2)_{L}$ transformations}}.}"

`` For each  [set of row and column labels] we shall also
\textbf{\emph{display a chunk of the map of the SM components}} of
[SO(10) multiplets] (i.e. the submultiplets with definite SM
quantum numbers) onto the defining basis states $H_{i}$,
$\Sigma_{ijklm}$, $\overline\Sigma_{ijklm}$, $\Phi_{ijkl}$ and
$\Psi_{ijk}$ (typically we present only the ``lowest'' relevant
permutation of indices and defer an interested reader to Appendix
  or to \cite{bmsv} for further details) \textbf{\emph{in
order to provide an information about the phase convention used in
derivation of the mass matrix under consideration}}. (Note that
for sake of simplicity we always choose our phase convention in
such a way there are no pending imaginary units in the mass
matrices.) For sake of a simple bookkeeping the top-left box of
each table shall indicate the full dimensionality of the sector
under consideration."

After passing through a list of the component combinations and
phase choices of more than 100  separate SM states the author
concludes :

``To conclude, in this section we have written down the mass
matrices for all the 592 bosonic degrees of freedom (up to gauge
transformations) of the Higgs sector of the next-to-minimal
supersymmetric SUSY SO(10) model."

This method is to be contrasted with our method for handling
SO(10) invariants\cite{ag1} which we briefly recapitulate in order
to make clear the difference in approach. We quote directly from
\cite{ag1}:

\subsection{Translation from SO(10) vector to PS labels }

 We have adopted the rule that
any submultiplet of an SO(10) field is always denoted by the
{\it{same}} symbol as its parent field, its identity being
established by the indices it carries or by supplementary indices,
if necessary. Our notation for indices is as follows : The indices
of the vector representation of SO(10) (sometimes also SO(2N))
 are denoted by $i,j =1..10 (2N)$. The {\it {real}}
vector index of the upper left block embedding (i.e. the embedding
specified by the breakup of the vector multiplet $10=6 + 4$) of
SO(6) in SO(10) are denoted $a,b=1,2..6$ and of the lower right
block embedding of SO(4) in SO(10) by
${\tilde{\alpha},\tilde\beta= 7,8,9,10}$. These indices are
complexified via a Unitary transformation and denoted by
$\hat{a},\hat{b}=\hat{1},\hat{2},\hat{3},\hat{4},\hat{5},\hat{6}
\equiv \overline{\mu},\overline{\mu}^{*}=
\bar{1},{\bar{1}^*,\bar{2},\bar{2}^*,\bar{3},\bar{3}^*}$ where
$\hat{1}\equiv \bar{1}, \hat{2}\equiv \bar{1}^* $ etc. Similarly
we denote the complexified versions of
${\tilde{\alpha},\tilde\beta}$ by $\hat{\alpha},\hat{\beta}=
\hat{7},\hat{8},\hat{9},\widehat{10}$. The indices of the doublet
of SU(2)$_L(\rm{SU(2)}_R$) are denoted
$\alpha,\beta=1,2$($\dot\alpha,\dot\beta=\dot{1},\dot{2})$.
Finally the index of the fundamental 4-plet of SU(4) is denoted by
a (lower) $\mu,\nu = 1,2,3,4$ and its upper-left block SU(3)
subgroup indices are $\bar\mu,\bar\nu = 1,2,3$. The corresponding
indices on the $4^{*}$ are carried as superscripts.

\subsection  {\ SO(6) $ \longleftrightarrow  $\ SU(4)}
{{\bf{\underline{Vector/Antisymmetric}}}:  \vspace{.1 cm}
\noindent The 6 dimensional vector representation of SO(6) denoted
by $V_a
 (a=1,2,..,6)$ transforms as
\begin{equation}
V'_a=(exp {{\frac{i}{2}}\omega ^{cd}J_{cd}})_{ab}V_b
\end{equation}
where the Hermitian generators $J_{cd}$ have the explicit form
\begin{equation}
(J_{cd})_{ef}=-i\delta_{c[e}\delta_{f]d}
\end{equation}
and thus satisfy the SO(6) algebra (square brackets around indices
denote antisymmetrization)
\begin{equation}
[J_{cd},J_{ef}]= i\delta_{e[c}J_{d]f}-i\delta_{f[c}J_{d]e}
\end{equation}
It is useful to introduce complex indices $\hat a, \hat b = {\hat
1} ...{\hat 6}  $ by the unitary change of basis \bea V_{\hat
a}=U_{\hat a {a}}V_a \;, \quad U=U_2 \times I_3\;, \quad
U_2={\frac{1}{\sqrt{2}}}
   \left[  \begin{array}{cc}
1 & i \\ 1 & -i
\end{array}\right]
\eea so that  $V_{a}W_{a}=V_{\hat a}W_{\hat a^*}$. The
decomposition of the fundamental 4-plet of  SU(4) w.r.t.
SU(3)$\times \rm{U(1)}_{B-L}$ is $4=(3,1/3) \oplus(1,-1)$. The
index for the 4 of SU(4) is denoted by $\mu=1,2,3,4$ while $\bar
\mu=1,2,3$ label its SU(3) subgroup.
In SU(4) labels,~the 6 of  SO(6) is the 2 index antisymmetric
$V_{\mu\nu}$ and decomposes as $6=V_{\bar\mu}(3,-2/3) \oplus
V_{{\bar\mu}^* }(\bar 3, 2/3)$ and we identify
$V_{\bar\mu4}=V_{\bar\mu}$,~$V_{\bar\mu\bar \nu}=\epsilon_{\bar\mu
\bar \nu\bar \lambda}V_{\bar\lambda^*}$. In other words, if one
defines $V_{\mu\nu}=\Theta_{\mu\nu}^{\hat a} V_{\hat a}$ with
$\Theta_{{\bar \mu} 4}^{\hat a}=\delta^{\hat a}_{\bar \mu},
 \Theta_{{\bar \mu}{\bar \nu}}^{\hat a} =
\epsilon_{\bar \mu \bar \nu \bar \lambda}\delta^{\hat a}_{\bar
\lambda^*}$,~then since $\Theta_{\mu\nu}^{\hat a}
\Theta_{\lambda\sigma}^ {\hat a*} \equiv
\epsilon_{\mu\nu\lambda\sigma}$ it follows that the translation of
SO(6) vector index contraction is (${\widetilde V}^{\mu\nu}
=(1/2)\epsilon^{\mu\nu\lambda\sigma} V_{\lambda \sigma})$
\bea
V_aW_a&=&{\frac{1}{4}}\epsilon^{\mu\nu\lambda\sigma}V_{\mu\nu}W_{\lambda\sigma}
\equiv {\frac{1}{2}}\widetilde V^{\mu\nu}W_{\mu\nu}\label{6vec}\\
{\rm{while}}~~~
V_aW_a^*&=&{1 \over 2} V_{\mu\nu}(W_{\mu\nu})^* \eea
Representations carrying vector indices $a,b  ...$ are then
translated by replacing by each vector index by an antisymmetrized
pair of SU(4) indices $\mu_1\nu_1, \mu_2\nu_2......$. For example
\bea A_{ab} B_{ab} &= &2^{-4} \epsilon^{\mu_1\mu_2\mu_3\mu_4}
\epsilon^{\nu_1\nu_2\nu_3\nu_4} A_{{\mu_1\mu_2},{\nu_1\nu_2}}
B_{{\mu_3\mu_4},{\nu_3\nu_4}}\\
{\rm{while}}~~~~~
A_{ab} B^*_{ab}& =& 2^{-2}  A_{{\mu_1\mu_2},{\nu_1\nu_2}}
B^*_{{\mu_1\mu_2},{\nu_1\nu_2}} \eea

\subsection{$\rm{SO(4)} \leftrightarrow \rm{SU(2)}_{L} \times \rm{SU(2)}_{R}$}
{{\bf{\underline{Vector/Bidoublet}}} \vspace{.1 cm}\\ We use early
greek indices ${ \ta, \tb =7,8,9,10}$ for the vector of SO(4)
corresponding to ${i,j=7....10}$ of the 10-plet of SO(10). The
Hermitian generators of SO(4) have the usual SO(2N) vector
representation form :
${(J_{\tilde\alpha\tilde\beta})_{\tilde\gamma\tilde
\delta}}=-i{\delta_{\tilde\alpha[\tilde\gamma}\delta_{\tilde\delta]\tilde
\beta}} $.\\ The group element is $R=exp {i \over
2}\omega^{\tilde\alpha\tilde\beta}J_ {\tilde\alpha\tilde\beta}$.
The generators of SO(4) separate neatly into self-dual and
anti-self- dual sets of 3, ${J_{\tilde
\alpha\tilde\beta}^{\pm}}={1 \over 2}{(J_{\tilde\alpha\tilde\beta}
\pm {\tilde J_{\tilde\alpha\tilde\beta})}}$. Then if
${\check\alpha,\check\beta=1,2,3}$ the generators and parameters
of the $SU(2)_{\pm}$ subgroups of SO(4) are defined to be \be
J_{\check\alpha}^{\pm}={1 \over
2}\epsilon_{\check\alpha\check\beta\check \gamma} J_{(\check\beta
+6)(\check\gamma +6)}^{\pm} ~ ;\quad
\omega_{\check\alpha}^{\pm}={1 \over 2}\epsilon_{{\check
\alpha}{\check \beta }{\check\gamma}}\omega_{{(\check\beta+6})
(\check\gamma +6) } \pm \omega_{(\check\alpha+6)10} \ee
The $SU(2)_{\pm}$ group elements are ${exp(i{\vec \omega}_{\pm}
\cdot \vec J^{\pm})}.$ The vector 4-plet of SO(4) is a bi-doublet
$(2,2)$ w.r.t. to ${SU(2)_{-} \otimes SU(2)_{+}}$.
We denote the indices of the doublet of ${SU(2)_L=SU(2)_{-}}$
${(SU(2)_R= SU(2)_{+})}$ by undotted early greek indices
${\alpha,\beta=1,2}$ {(dotted early greek indices
$\dot\alpha,\dot\beta=\dot{1},\dot{2}$)}. Then one has \bea
V_{\hat7}&=& V_{\bar 4}= {{(V_7+iV_8)} \over \sqrt2}=V_{2\dot2}
~,~ V_{\hat9}= V_{\bar 5}=
{{(V_9+iV_{10})}\over \sqrt2}=V_{1\dot2}\\
V_{\hat 8}&=& V_{\bar4^*}={{(V_7-iV_8)} \over
\sqrt2}=-V_{1\dot1}~,
 ~V_{\widehat{10}}\equiv V_{\hat 0}= V_{\bar5^*}={{(V_9-iV_{10})}\over \sqrt2}=V_{2\dot1}
\label{bidoub} \eea ${SU(2)_L {(SU(2)_R)}}$ indices are raised and
lowered with ${\epsilon^ {\alpha\beta},\epsilon_{\alpha\beta}}$
${(\epsilon^{\dot\alpha\dot\beta},
\epsilon_{\dot\alpha\dot\beta})}$ with
${\epsilon^{12}=+\epsilon_{21}=1}$ etc. The SO(4) vector index
contraction translates as \bea V_{\tilde\alpha}W_{\tilde\alpha}
&=& -{V_{\alpha\dot\alpha}W_{\beta\dot\beta}
\epsilon^{\alpha\beta}\epsilon^{\dot\alpha\dot\beta}}
=-{V^{\alpha\dot\alpha}W_{\alpha\dot\alpha}}\\ {\rm{While}}~~~~~~~
V_{\tilde\alpha}W_{\tilde\alpha}^*
&=&{V_{\alpha\dot\alpha}W^{*}_{\alpha\dot\alpha}}\label{4vec} \eea

\textbf{\emph{Thus the above rules enable the decomposition of any
SO(10) tensor invariant not involving spinors into PS invariants
(and then trivially into SM invariants)}} by decomposing the SU(4)
quartet w.r.t SU(3)$\times \rm{U(1)}_{B-L}$ as $4=(3,1/3)
\oplus(1,-1)$ and  relating the SM hypercharge to $T_{3R},B-L$ in
the standard way\cite{mohamarsh} : $Y/2= T_{3R} + (B-L)/2 $.

It remains to quote the same for the spinor indices :

The Clifford algebra of SO(2N) acts on a $2^N$ dimensional space
which is given the convenient basis of eigenvectors
${|\epsilon=\pm 1>}$ of $\tau_3$: \be
{|\epsilon_1,.......\epsilon_n>}={|\epsilon_1>}\otimes......\otimes{|\epsilon_n>}
\ee In this basis $\gamma_F=\prod_{i=1}^{n}\epsilon_i$. So the
basis spinors of SO(2N) decompose into odd and even subspaces
w.r.t. $\gamma_F$. \be 2^n=2_{+}^{n-1}+2_{-}^{n-1} \ee

\subsection{SO(6) Spinors}
The $4(\psi_\mu)$ and $\bar 4(\widehat{\psi}^\mu)$ of SU(4) may be
consistently identified with the $4_{-},4_{+}$ chiral spinor
multiplets of SO(6) by identifying components ${\psi_\mu}$ of the
4 with the coefficients of the states
$|\epsilon_1\epsilon_2\epsilon_3>_{-}$ in $4_{-}=|\psi>_{-}$ as
\be |\psi>_{-}=\psi_{1}|-++> +~\psi_{2}|+-+> +~\psi_3|++->
+~\psi_4|---> \ee and also ${\widehat\psi^\mu}$ in the
$4_{+}=|\widehat\psi>_{+}$ as \be
|\widehat\psi>_{+}=-\widehat\psi^{1}|+--> +~\widehat\psi^{2}|-+->
- ~\widehat\psi^{3}|--+> +~\widehat\psi^{4}|+++> \ee The reason
for the extra minus signs is that then the charge conjugation
matrix $C_2^{(3)}$ correctly combines the $4,\bar 4$ components in
the $2^3$-plet spinors of SO(6) to make SU(4) singlets and
covariants . For example (we take $\psi,\chi$ to be non-chiral
$8=4_{+}+4_{-}$ spinors to preserve generality)

In this basis one has in the 8 dimensional spinor rep. of SO(6)

$$ exp({{i\omega^{ab}J_{ab}}\over 2}) = Diag {\big (}
exp({{i\theta^A \lambda^A}\over 2}) , exp({{-i\theta^A
\lambda^{A*}}\over 2}){\big )} $$

  One finds
the following useful identities hold \be
\begin{array}{cl}
\psi^{T}{C}_{2}^{(3)}\chi&=\psi_{\mu}{\widehat\chi}^{\mu}+{\widehat\psi}^{\mu}
\chi_{\mu}=\psi.\widehat\chi+\widehat\psi.\chi\\
\psi^{T}{C}_{2}^{(3)}\gamma_{\mu\nu}\chi&=\sqrt{2}{(-\psi_{[\mu}\chi_{\nu]}+
{\widehat\psi}^{\lambda}{\widehat\chi}^{\sigma}\epsilon_{\mu\nu\lambda\sigma})}\\
\psi^{T}{C}_{2}^{(3)}\gamma_{\mu\nu}\gamma_{\lambda\sigma}\chi&=-2{\{\widehat\psi^
{\theta}\chi_{[\lambda}\epsilon_{\sigma]\mu\nu\theta}+\psi_{[\mu}\epsilon_{\nu]
\lambda\sigma\theta}\widehat\chi^{\theta}\}}\\
\psi^{T}{C}^{(3)}_{2}\gamma_{\mu\nu}\gamma_{\lambda\sigma}\gamma_{\theta\delta}
\chi&={(\sqrt{2})^3}{\{\psi_{[\mu}\epsilon_{\nu]\lambda\sigma[\theta}\chi
_{\delta]}+{\widehat\psi}^{\omega}\widehat\chi^{\rho}\epsilon_{\omega\mu\nu
[\theta}\epsilon _{\delta]\rho\lambda\sigma}\}}
\end{array}\label{so6spin}
\ee The results when $\psi^{T}C^{(3)}_{2}\rightarrow
\psi^{\dagger}$ are obtained by the replacements
$\psi_\mu\rightarrow \widehat{\psi}^{\mu*}$  and
$\widehat{\psi}^{\mu}\rightarrow\psi^{*}_{\mu}$ on the R.H.S of
all the identities in (\ref{so6spin}). The square root factors
arise because the antisymmetric pair labels for the gamma matrices
correspond to complex indices ${\hat a}, {\hat b}$. Note that one
does not need the identities for more than 3 gamma matrices. See
the appendix of \cite{ag1} for useful translations of SO(6)
spinor-tensor invariants calculable from these identities .
\subsection {SO(4) Spinors}
\noindent In the case of SO(4) the spinor representation is 4
dimensional and splits into $2_+\oplus 2_-$. It is not hard to see
that with the definitions adopted for the generators of
$SU(2)_{\pm}$ the  chiral spinors $2_{\pm}$ may be identified with
the doublets ${\psi_{\alpha},\psi_{\dot\alpha}}$ of
$SU(2)_{-}=SU(2)_L$ and $SU(2)_{+}=SU(2)_R$ as \be
|2>_{-}=|\psi>_{-}=\psi_{1}|+-> +~\psi_{2}|-+> ,\quad
|2>_{+}=|\psi>_{+}=\psi_{\dot{1}}|++> -~\psi_{\dot{2}}|--> \ee As
in the SO(6) case one transforms to the unitary basis where
$4=2_{+}\oplus 2_{-}$ has components $(\psi_{\alpha},\psi_{\da})$.
Then in that basis \be C_{2}=\left(\begin{array}{cc}
\epsilon^{\alpha\beta} & 0_{2}\\ 0_{2} & -\epsilon^{\da\db}
\end{array}\right)~,~C_{1}=-\left(\begin{array}{cc}
\epsilon^{\alpha\beta} & 0_{2}\\ 0_{2} &
\epsilon^{\da\db}\end{array}\right)~,
~[\gamma_{\rho\dot\rho}]=\sqrt{2}\left(\begin{array}{cc} 0_{2} &
\epsilon_{\rho\alpha}\delta^{\db}_{\dot\rho}
\\\epsilon_{\dot\rho\da}\delta^{\beta}_{\rho} & 0_{2}
\end{array}\right)
\ee

 The following expressions for spinor covariants then follow
\be
\begin{array}{cl}
{\psi}^{T}C_{2}^{(2)}\chi&=\psi^{\dot\alpha}\chi_{\dot\alpha}-\psi^{\alpha}
\chi_{\alpha}\\
{\psi}^{T}C_{1}^{(2)}\chi&=\psi^{\dot\alpha}\chi_{\dot\alpha}+\psi^{\alpha}
\chi_{\alpha}\\
\psi^{T}{C}_{2}^{(2)}\gamma_{\alpha\dot\alpha}\chi&=\sqrt{2}{(
\psi_{\dot
\alpha}\chi_{\alpha}-\psi_{\alpha} \chi_{\dot\alpha})}\\
\psi^{T}{C}_{1}^{(2)}\gamma_{\alpha\dot\alpha}\chi&=\sqrt{2}{(
\psi_{\dot
\alpha}\chi_{\alpha}+\psi_{\alpha} \chi_{\dot\alpha})}\\
\psi^{T}{C}_{2}^{(2)}\gamma_{\alpha\dot\alpha}\gamma_{\beta\dot\beta}\chi&=
2{\epsilon_{\dot\alpha\dot\beta}\psi_{\alpha}\chi_{\beta}-2\epsilon_{\alpha
\beta} \psi_{\dot\alpha} \chi_{\dot\beta}}\\
%
\psi^{T}{C}_{1}^{(2)}\gamma_{\alpha\dot\alpha}\gamma_{\beta\dot\beta}\chi&=
{-2\epsilon_{\dot\alpha\dot\beta}\psi_{\alpha}\chi_{\beta}-2\epsilon_{\alpha
\beta} \psi_{\dot\alpha} \chi_{\dot\beta}}
\end{array}\label{so4spin}
\ee

Furthermore \be
\begin{array}{cl}
\psi^{\dagger}\chi&=\psi^{*}_{\da}\chi_{\da}+\psi^{*}_{\alpha}\chi_{\alpha}\\
\psi^{\dagger}\gamma_{\alpha\da}\chi&=-{\sqrt{2}}(\psi^{\alpha*}\chi_{\da}+\psi^{\da*}\chi_{\alpha})\\
\psi^{\dagger}\gamma_{\alpha\da}\gamma_{\beta\db}\chi&=2\epsilon_{\da\db}
\psi^{\alpha*}\chi_{\beta}+2\epsilon_{\alpha\beta}\psi^{\da*}\chi_{\db}
\label{so4spin2}
\end{array}
\ee Note that these can be obtained from the corresponding
identities involving $C^{(2)}_{1}$ by the replacements
$\psi^{\da}\rightarrow \psi^{*}_{\da},\psi^{\alpha}\rightarrow
\psi^{*}_{\alpha}$ or from the $C_{2}$ identities by
$\psi^{\da}\rightarrow
\psi_{\da}^{*},\psi^{\alpha}\rightarrow-\psi_{\alpha}^{*}$.
\subsection {SO(10) Spinors}
The spinor representation of SO(10) is $2^5$ dimensional and
 splits into chiral eigenstates with $\gamma_F=\pm 1$ as
\bea
2^5&=&2^{4}_{+}+2^{4}_{-}=16_{+}+16_{-}\\
16&=&16_{+}=(4_{+},2_{+})+(4_{-},2_{-})=(\overline{4},1,2)+(4,2,1) \\
\overline{16}&=&16_{-}=(4_{+},2_{-})+(4_{-},2_{+})=(\overline{4},2,1)+
(4,1,2) \eea Where the first equality follows from the definition
of $\gamma_F$ and the second from the SO(6) to SU(4) and SO(4) to
$SU(2)_{L} \times SU(2)_{R}$ translations: $4_-=4,2_+=2_{R}
,2_-=2_{L}$. Thus we see that the SU(4) and $SU(2)_L \times
SU(2)_R$ properties of the submultiplets within the
$16,\overline{16}$ are strictly correlated. Use of the SO(6) and
SO(4) spinor covariant identities allows fast construction of
SO(10) spinor invariants. For example , \be \psi^{T}C_{2}^{(5)}
\gamma_{\mu\nu}^{(5)}\chi = \psi^{T}(C_2^{(3)} \times
C_{1}^{(2)})(\gamma_{\mu\nu}^{(3)} \times \tau_3 \times
\tau_3)\chi =\psi^T(C_2^{(3)}\gamma_{\mu\nu}^{(3)}\times
C_2^{(2)})\chi \ee Next one uses the identities
(\ref{so6spin},\ref{so4spin}) in parallel , keeping in mind that
in the 16-plet the dotted ($\rm{SU(2)}_R $) spinors are always
$\bar 4$-plets of SU(4) and the undotted ones are 4-plets and vice
versa for $\overline{16}$ . When $\psi,\chi$ are both 16-plets one
immediately reads off the result \be
\psi^{T}C_{2}^{(5)}\gamma_{\mu\nu}^{(5)}\chi=\sqrt
{2}{(\psi^{\alpha}_{[\mu}\chi_{{\nu]}\alpha}+
{\widehat\psi}^{\lambda{\dot\alpha}}{\widehat\chi}^{\sigma}_{\dot\alpha}
\epsilon_{\mu\nu\lambda\sigma})} \ee

In addition to the above rules for the decomposition of the two
fundamental irreps of $\mathbf{Spin(10)}$ we also gave \cite{ag1}
extensive tables of decompositions of SO(10) cubic invariants e.g
$\mathbf{16.16.(10\oplus120\oplus\oot)}$ and even the matter
kinetic terms that are directly usable in the SO(10)
superpotential. Finally the standard PS embedding of a matter
fermion generation in the 16- plet completes the specification of
all normalizations and phases from the SO(10) down to the SM.

\bea  (4,2,1) = (Q_{\alpha},L_{\alpha} ) \qquad \qquad
({\overline{4}},1,2) =
 ({\overline{Q}}_{\alpha},{\overline{L}}_{\alpha}) \eea

with
\begin{equation}
Q=\left({\begin{array}{c}U\\D\end{array}}\right)
 \quad L=\left({\begin{array}{c} \nu\\e\end{array}}\right) \quad
{\overline Q}=\left({\begin{array}{c} {\bar d}\\{\bar
u}\end{array}}\right) \quad
{\overline{L}}=\left({\begin{array}{c}{\bar e}\\{\bar
\nu}\end{array}}\right) \quad
\end{equation}

 Note that in \cite{ag1} and thereafter we only perform unitary basis
transformations on the fields  and thus always maintain unit norm
as defined by unit coefficient of  canonical kinetic terms in the
Lagrangian.

We conclude with a quote from\cite{ag2} which summarizes our
consistent position on the relative merits of the component wise
and systematic  decomposition via maximal sub-group  approaches  :
``We emphasize that our method allows computation,
 not only of spectra but also of the  couplings  of all  the multiplets  in the
theory (whether they are renormalizable or  heavy-exchange induced
effective couplings) without any ambiguity. Moreover our results
are obtained by  an analytic tensorial reprocessing of labels of
fields in the Lagrangian.  This approach might thus
 find  preferment with field theorists
 in comparison with the more restricted capabilities
   of the approach of \cite{heme}, which, so far, has not proved capable
 of generating all the Clebsches of the SO(10)
theory and which  relies on an explicit multiplet
 representative and computer based approach
which is  tedious to connect to the    unitary group tensor
methods so familiar to particle theorists."

\subsection{Fermion Yukawa couplings}

Since it is so simple let us summarize our prescription for
determining the MSSM couplings in terms of GUT ones. When we
rewrite the equations of \cite{malinsky} in our notation the
non-holomorphic connection problem will become all too obvious.

We call the (6 pairs of)  EW type doublets contained in the SO(10)
fields\cite{blmdm,nmsgut} $ \{ \bar{h}_i[1,2,-1]\oplus h_i[1,2,1];
i=1...6\}  $ and their mass matrix due to mass terms in the
superpotential ${\cal{H}}(x,\lambda)$. Here we have exhibited the
(holomorphic) dependence of the mass matrix on the generic
vevs($x$) and superpotential couplings($\lambda$). The mass
eigenstates of ${\cal{H}}$ are called $ \{ \bar{H}_i[1,2,-1]\oplus
H_i[1,2,1]; i=1...6\} $. A bi-unitary transformation connects the
two sets : \be h^{(i)} = U_{ij} H^{(j)} \qquad ;\qquad {\bar
h}^{(i)} = { \overline U}_{ij} {\overline  H}^{(j)} \ee
 The columns of $U(\bar U) $ are the unit normalized right eigenvectors of
${\cal H}^\dagger {\cal H}({\cal H}^*{\cal H}^T)$.   When a fine
tuning condition $Det({\cal H})=0$ is
imposed\cite{abmsv,bmsv,ag1,ag2}  one pair of doublets remains
light i.e , massless on the scale $M_X$ which we call
$\{\bar{H}^{(1)},H^{(1)}\}$. Then it follows that
    \bea \Lambda_H &=&Diag(m_H^{(1)},m_H^{(2)},....)  =
 {\overline U}^T {\cal H} U\nnu
 W &=&  \bar{h}^T {\cal{H}} h +... =\bar{H}^T \Lambda_H H +...
\nonumber \eea where the first eigenvalue $m^{(1)}_H=0$ and the
rest can always be made positive by a choice of the phases of the
eigenvectors. Then the passage to the  renormalizable  effective
MSSM is simple indeed : simply set all the Superheavy Higgs
doublets $ \{ \bar{H}_i[1,2,-1]\oplus H_i[1,2,1]; i=2...6\} $ to
zero in all SO(10) invariants involving them ! In other words \bea
h^{(i)} \rightarrow  \alpha_i H^{(1)}=   \alpha_i H\quad ; \quad
{\bar h}^{(i)} \rightarrow  {\bar\alpha}_i {\overline H}^{(1)}={
\bar\alpha}_i {\overline  H}\label{superhhbar}\eea where the
numbers $ \alpha_i = U_{i1} , {\bar\alpha}_i =\bar U_{i1}$, which
for obvious reasons we call \emph{Higgs fractions},   are the
critical information which is to be extracted by diagonalizing
${\cal{H}} $. It is clear that   the un-normalized Higgs fractions
$\hat\alpha_i$(${\hat{\bar{\alpha}}}_i$)\cite{nmsgut}  are the
right and left null eigenvectors of ${\cal H}$ and  are
holomorphic in the vevs and couplings.

The inverse transformations are obviously $H=U^{\dagger}
{h},{\overline H}= {\overline  U}^{\dagger} \bar{h}$ so that in
particular \bea H^{(1)}\equiv H &=&  (U^{\dagger})_{1j} h^{(j)} =
(U^*)_{j1} h^{(j)}  =\alpha_j^* h^{(j)} \nnu  {\overline
H}^{(1)}\equiv {\overline H} &=& ( {\overline U}^{\dagger})_{1j}
\bar{h}^{(j)}  = ( {\overline  U}^*)_{j1} \bar{h}^{(j)}
=\bar{\alpha}_j^* \bar{h}^{(j)}  \label{lightHHbar} \eea So that
the coefficients of the substitution rule (\ref{superhhbar}) are
the unconjugated Higgs fractions while those in  equation
(\ref{lightHHbar}) for the light Higgs in terms of the   GUT
doublets are conjugate. Note that the relation between SO(10)
chiral multiplets and MSSM ones is holomorphic as it should be.
Thus all relevant information is contained in the Higgs
fractions$\{\alpha_i,\bar{\alpha}_i ; i=1,6\}$ alone.

Let us now turn to the corresponding equations in \cite{malinsky}
and explain why they are incorrect. We shall quote
 the necessary  equations directly from  that paper but also repeat them in our notation
 {  {but with primes  on corresponding quantities}} to make
the correspondence perfectly clear.   Corresponding to the
equation (\ref{lightHHbar})  \cite{malinsky} has :\bea
h_{u}&\propto&
w^{u}_{10}H^{u}+w^{u}_{\overline{126}}\overline{\Sigma}^{u}+w^{u}_{{126}}
{\Sigma}^{u}+w^{u}_{210}\Phi^{u}+w^{u(1)}_{120}\Psi^{u}_{(1)}+w^{u(2)}_{120}\Psi^{u}_{(2)}\nn\\
h_{d}&\propto&
w^{d}_{10}H^{d}+w^{d}_{{126}}{\Sigma}^{d}+w^{d}_{\overline{126}}
\overline{\Sigma}^{d}+w^{d}_{210}\Phi^{d}+w^{d(1)}_{120}\Psi^{d}_{(1)}+w^{d(2)}_{120}\Psi^{d}_{(2)}
\label{doubletweights} \eea which in our notation would read \bea
H'= \alpha_j'  h^{(j)'}  \qquad ;
 \qquad {\overline H}'= \bar{\alpha}_j'  \bar{h}^{(j)'}\eea
 Comparing with (\ref{lightHHbar}) the coefficient functions
$w^{d,u}_i=\alpha_i',\bar\alpha_i'$ which are given as
unconjugated in \cite{malinsky}  are seen to lack a conjugation
and  should be rather $\alpha_i^*,\bar\alpha_i^*$.    Conversely
when we examine the correspondents of the substitution
rule(\ref{superhhbar}) we find that they are the equations (47,48)
of \cite{malinsky} : `` Let us {\it define} the projections of the
electroweak doublet VEVs onto the neutral components of the
defining basis doublets $H^{u,d}$, $\overline\Sigma^{u,d}$,
$\Psi^{u,d}_{(1)}$ and $\Psi^{u,d}_{(2)}$ as follows: \bea
\vev{H^{u}}\equiv u_{10}^{u},\;\; \vev{\overline \Sigma^{u}}
\equiv u_{\overline{126}}^{u},\;\;\vev{\Psi^{u}_{(1)}}\equiv
u_{120}^{u(1)},
\;\;\vev{\Psi^{u}_{(2)}}\equiv u_{120}^{u(2)},\nn\\
\vev{H^{d}}\equiv u_{10}^{d},\;\;\vev{\overline \Sigma^{d}}\equiv
u_{\overline{126}}^{d},\;\;\vev{\Psi^{d}_{(1)}}\equiv
u_{120}^{d(1)},\;\;\vev{\Psi^{d}_{(2)}}\equiv u_{120}^{d(2)}.\;
\label{ewvevs} \eea The main virtue of this definition is that
these factors {\it are} indeed simple functions of the
decomposition weights in (\ref{doubletweights}) and the VEVs
$v_{u}$ and $v_{d}$ of the MSSM light Higgs doublets
($\vev{h_{u,d}}\equiv v_{u,d}$), namely : \be\label{projwts}
u^{u,d}_{10}=(w^{u,d}_{10}v_{u,d})^{*},\;\;
u^{u,d}_{\overline{126}}=(w^{u,d}_{\overline{126}}v_{u,d})^{*},\;\;
u^{u,d(1)}_{120}=(w^{u,d(1)}_{120}v_{u,d})^{*}\;\text{ and }\;
u^{u,d(2)}_{120}=(w^{u,d(2)}_{120}v_{u,d})^{*} \ee ".

 The
equation(\ref{ewvevs}) is simply a name for vevs so it is not
clear what significance the emphasis on  \emph{define} has for the
author; one could stay with lrangles to indicate vevs just as
well.  We found the next equation(\ref{projwts}) had an error :
\textbf{\emph{ the field relations were anti-holomorphic}} and the
Higgs fractions were conjugated but in our substitution rule they
are not. Thus further comparison was pointless till the
discrepancies are corrected.

\section{Discussion}

In this comment we have taken the trouble to deconstruct the
calculation of \cite{malinsky} to make clear that there are
  defects in the method of calculation.  In our view use of group theoretic methods adequate
  to conveniently, systematically and unambiguously encode  the
  complexities of the embeddings in SO(10) and the decomposition    w.r.t. the SM subgroup
    in a compact and generative(i.e tensor index rule based rather than adhoc)
      are essential. Such methods  should not  be based on voluminous
  lists of arbitrary phase choices but   on generative
  rules that can be repeated applied with confidence that they
  will be consistent globally.  We  provided one such framework in\cite{ag1}. Another
  might perhaps be based on \cite{mohsak,nathsyed}. It would
  certainly be highly interesting to cross check the conventions,
  phases and normalizations between the two different maximal
  subgroup methods. The explicit state based methods used
  in\cite{malinsky} and its antecedent papers are, in our view,
  useful only for a limited purpose of checking magnitudes and
  will never serve as an efficient basis for computation in SO(10)
  GUTs in all their complex interactive glory. The arbitrary phase
  conventions used are all but impossible to check between the
  multiple computations that now exist\cite{nmsgut,bmsv,fuku,malinsky}.
   Thus they   create and  amplify
  Babel on the  SO(10) `tower' (or PetaPlex) construction site
    where we collectively labour,
  rather  than enable its speedy erection.

\section*{Acknowledgments}

I acknowledge  correspondence  with M. Malinsky which led to the
changes made in Version 2.


\begin{thebibliography}{1}


\bibitem{nmsgut}
  C.~S.~Aulakh and S.~K.~Garg,
 ``\emph{The new minimal supersymmetric GUT }'',
  arXiv:hep-ph/0612021.
\bibitem{malinsky}
  M.~Malinsky,
 ``Higgs sector of the next-to-minimal renormalizable SUSY SO(10),''
  arXiv:0807.0591 [hep-ph].
\bibitem{ag2}
  C.~S.~Aulakh and A.~Girdhar,
    \emph{SO(10) MSGUT: Spectra, couplings and threshold effects},
    Nucl.\ Phys.\ B {\bf 711}, 275 (2005).


\bibitem{ag1} C.S.Aulakh and A. Girdhar,
  {\it{SO(10) a la Pati-Salam }},
  hep-ph/0204097;v1 April 2002; v2 August 2003;
 v4, 9 February, 2004;
   Int.\ J.\ Mod.\ Phys.\ A {\bf 20}, 865 (2005)




\bibitem{mslrms}
  C.~S.~Aulakh, K.~Benakli and G.~Senjanovic,
  ``\emph{Reconciling supersymmetry and left-right symmetry},''
  Phys.\ Rev.\ Lett.\  {\bf 79} (1997) 2188
  [arXiv:hep-ph/9703434];
  C.~S.~Aulakh, A.~Melfo and G.~Senjanovic,
  ``\emph{Minimal supersymmetric left-right model},''
  Phys.\ Rev.\  D {\bf 57} (1998) 4174
  [arXiv:hep-ph/9707256];
  C.~S.~Aulakh, A.~Melfo, A.~Rasin and G.~Senjanovic,
 ``\emph{See-saw and supersymmetry or exact R-parity},''
  Phys.\ Lett.\  B {\bf 459} (1999) 557
  [arXiv:hep-ph/9902409];
  C.~S.~Aulakh, A.~Melfo, A.~Rasin and G.~Senjanovic,
  ``\emph{Supersymmetry and large scale left-right symmetry},''
  Phys.\ Rev.\  D {\bf 58} (1998) 115007
  [arXiv:hep-ph/9712551];
  C.~S.~Aulakh, B.~Bajc, A.~Melfo, A.~Rasin and G.~Senjanovic,
 ``\emph{Intermediate scales in supersymmetric GUTs: The survival of the  fittest},''
  Phys.\ Lett.\  B {\bf 460} (1999) 325
  [arXiv:hep-ph/9904352].

\bibitem{superK}Y. Fukuda et al(Super-Kamiokande Collaboration), Phys. Rev. Lett. {\bf 82}, 1810 (1999),
 Phys. Rev. Lett. {\bf 82}, 2430 (1999).

\bibitem{genealog}
  C.~S.~Aulakh,
  ``\emph{GUT genealogies for SUSY seesaw Higgs},''
  arXiv:hep-ph/0204098, Published in *Dubna 2001, Supersymmetry and unification of
fundamental interactions* 242-244( Proceedings of 9th
International Conference on   Supersymmetry and Unification of
Fundamental Interactions (SUSY01), Dubna, Russia, 11-17 Jun 2001).
.

\bibitem{abmrs01}
  C.~S.~Aulakh, B.~Bajc, A.~Melfo, A.~Rasin and G.~Senjanovic,
  ``\emph{SO(10) theory of R-parity and neutrino mass},''
  Nucl.\ Phys.\  B {\bf 597} (2001) 89
  [arXiv:hep-ph/0004031].

\bibitem{aulmoh} C.S.~Aulakh and R.N.~Mohapatra, CCNY-HEP-82-4 April 1982,
  CCNY-HEP-82-4-REV,  Jun 1982 , Phys. Rev. {\bf D28}, 217 (1983).
\bibitem{ckn} T.E. Clark, T.K.Kuo, and N.Nakagawa, Phys. lett. {\bf{115B}}, 26(1982).
\bibitem{nathsyed}
  P.~Nath and R.~M.~Syed,
  ``\emph{Complete cubic and quartic couplings of 16 and 16-bar in SO(10)
  unification},''
  Nucl.\ Phys.\  B {\bf 618}, 138 (2001)
  [arXiv:hep-th/0109116];
  P.~Nath and R.~M.~Syed,
  ``\emph{Analysis of couplings with large tensor representations in SO(2N) and
   proton decay},''
  Phys.\ Lett.\  B {\bf 506}, 68 (2001)
  [Erratum-ibid.\  B {\bf 508}, 216 (2001)]
  [arXiv:hep-ph/0103165].


\bibitem{abmsv}
  C.~S.~Aulakh, B.~Bajc, A.~Melfo, G.~Senjanovic and F.~Vissani,
  ``\emph{The minimal supersymmetric grand unified theory},''
  Phys.\ Lett.\  B {\bf 588} (2004) 196
  [arXiv:hep-ph/0306242].
\bibitem{bmsv}
B.~Bajc, A.~Melfo, G.~Senjanovic and F.~Vissani,
  \emph{The minimal supersymmetric grand unified theory. I: Symmetry breaking and
 the particle spectrum},
Phys.\ Rev.\ D {\bf 70}, 035007 (2004) [arXiv:hep-ph/0402122].

\bibitem{fuku}
   T.~Fukuyama, A.~Ilakovac, T.~Kikuchi, S.~Meljanac and N.~Okada,
  arXiv:hep-ph/0401213v1.,v2;
  T.~Fukuyama, A.~Ilakovac, T.~Kikuchi, S.~Meljanac and N.~Okada,
  J.\ Math.\ Phys.\  {\bf 46} (2005) 033505
  [arXiv:hep-ph/0405300].


\bibitem{heme} X.G. He and S. Meljanac, Phys. Rev. {\bf{D41}}, 1620 (1990).



\bibitem{consistency}
  C.~S.~Aulakh,
 ``\emph{On the consistency of MSGUT spectra},''
  Phys.\ Rev.\  D {\bf 72} (2005) 051702
  [arXiv:hep-ph/0501025].





\bibitem{allferm}   K.~Y.~Oda, E.~Takasugi, M.~Tanaka and
M.~Yoshimura,
  ``\emph{Unified explanation of quark and lepton masses and mixings in the
  supersymmetric SO(10) model},''
  Phys.\ Rev.\ D {\bf 59}, 055001 (1999)
  [arXiv:hep-ph/9808241]; K.~Matsuda, Y.~Koide, T.~Fukuyama and H.~Nishiura,
   ``\emph{How far can the SO(10) two Higgs model describe the observed neutrino
   masses and mixings?},''
  Phys.\ Rev.\ D {\bf 65}, 033008 (2002)
  [Erratum-ibid.\ D {\bf 65}, 079904 (2002)]
  [arXiv:hep-ph/0108202] ;
  K.~Matsuda, Y.~Koide and T.~Fukuyama,
   ``Can the SO(10) model with two Higgs doublets reproduce the observed
  fermion masses?,''
  Phys.\ Rev.\ D {\bf 64}, 053015 (2001)
  [arXiv:hep-ph/0010026].
  N.~Oshimo,
 ; Phys.\ Rev.\ D {\bf 66}, 095010 (2002)
  [arXiv:hep-ph/0206239];
   N.~Oshimo,
  Nucl.\ Phys.\ B {\bf 668}, 258 (2003)
  [arXiv:hep-ph/0305166]; B.~Bajc, G.~Senjanovic and F.~Vissani,
 ``\emph{b - tau unification and large atmospheric mixing: A case for non-canonical
 see-saw},''
Phys.\ Rev.\ Lett.\  {\bf 90} (2003) 051802
[arXiv:hep-ph/0210207];
 H.~S.~Goh, R.~N.~Mohapatra and S.~P.~Ng,
   ``\emph{Minimal SUSY SO(10), b tau unification and large neutrino mixings},''
  Phys.\ Lett.\ B {\bf 570}, 215 (2003)  [arXiv:hep-ph/0303055].
   H.~S.~Goh, R.~N.~Mohapatra and S.~P.~Ng,
   ``\emph{Minimal SUSY SO(10) model and predictions for neutrino mixings and
   leptonic CP violation},''
  Phys.\ Rev.\ D {\bf 68}, 115008 (2003)
  [arXiv:hep-ph/0308197].
  H.~S.~Goh, R.~N.~Mohapatra and S.~Nasri,
  Phys.\ Rev.\ D {\bf 70} (2004) 075022
  [arXiv:hep-ph/0408139];
S.~Bertolini, M.~Frigerio and M.~Malinsky,
Phys.\ Rev.\ D {\bf 70}, 095002 (2004)
[arXiv:hep-ph/0406117]; 
   S.~Bertolini and M.~Malinsky,
  arXiv:hep-ph/0504241 ;
     K.~S.~Babu and C.~Macesanu,
  arXiv:hep-ph/0505200.
  B.~Bajc, G.~Senjanovic and F.~Vissani,
  Phys.\ Rev.\ D {\bf 70}, 093002 (2004)
  [arXiv:hep-ph/0402140];
  B.~Bajc, G.~Senjanovic and F.~Vissani,
  arXiv:hep-ph/0110310;K.~S.~Babu and C.~Macesanu,
  Phys.\ Rev.\ D {\bf 72}, 115003 (2005)
  [arXiv:hep-ph/0505200].



\bibitem{bmsv0511}  B.~Bajc, A.~Melfo,
G.~Senjanovic, and F.~Vissani,
\newblock (2005), hep-ph/0511352.




\bibitem{gmblm} C.~S.~Aulakh, \emph{From germ to bloom},
 arXiv:hep-ph/0506291.
\bibitem{blmdm}  C.~S.~Aulakh and S.~K.~Garg,
 \emph{MSGUT: From bloom to doom},
  Nucl.\ Phys.\ B {\bf 757}, 47 (2006)
  [arXiv:hep-ph/0512224].


\bibitem{pinmsgut}
  C.~S.~Aulakh,
   ``\emph{Pinning down the New Minimal Supersymmetric GUT},''
  Phys.\ Lett.\  B {\bf 661}, 196 (2008)
  [arXiv:0710.3945 [hep-ph]].



\bibitem{bertmal}
  S.~Bertolini, T.~Schwetz and M.~Malinsky,
  Phys.\ Rev.\  D {\bf 73} (2006) 115012
  [arXiv:hep-ph/0605006].


\bibitem{core}   C.~S.~Aulakh,
 \emph{Fermion mass hierarchy in the Nu MSGUT. I: The real core},
  arXiv:hep-ph/0602132 ;

 \bibitem{msgreb}  C.~S.~Aulakh, \emph{MSGUT Reborn ?} arXiv:hep-ph/0607252

\bibitem{precthresh}
  C.~S.~Aulakh and S.~K.~Garg,
   ``\emph{Correcting $\alpha_3(M_Z)$ in the NMSGUT},''
  arXiv:0710.4018 [hep-ph].

\bibitem{mohamarsh}
  R.~E.~Marshak and R.~N.~Mohapatra,
  ``\emph{Quark - Lepton Symmetry And B-L As The U(1) Generator Of The Electroweak
   Symmetry Group},''
  Phys.\ Lett.\  B {\bf 91} (1980) 222.

\bibitem{mohsak}
  R.~N.~Mohapatra and B.~Sakita,
  ``\emph{SO(2N) grand unification in an SU(N) basis},''
  Phys.\ Rev.\  D {\bf 21} (1980) 1062.


\bibitem{nmsgutII}
  C.~S.~Aulakh and S.~K.~Garg,
 ``\emph{Nmsgut II: Pinning the Nmsgut@LHC}'',
  arXiv:0807.0917 [hep-ph].
\bibitem{nmsgut@cpt@ictp}
  C.~S.~Aulakh,
 `` \emph{Pinning the NMSGUT@LHC},'', Invited Talk, at CPT@ICTP, Workshop on the Origin of P, CP and T
 Violations, ICTP, Trieste, July2-5, 2008.


\end{thebibliography}

\end{document}